# Comparative Analysis of AES, Blowfish, Twofish, Salsa20, and ChaCha20 for Image Encryption

Rebwar Khalid Muhammed[a], Ribwar Rashid Aziz[b], Alla Ahmad Hassan[b], Aso Mohammed Aladdin[c], Shaida Jumaah Saydah[d], Tarik Ahmed. Rashid[e], Bryar Ahmad Hassan[c, e]

[a]Network Department, Computer Science Institute, Sulaimani Polytechnic University, Sulaymaniyah, Iraq
[b]Database Department, Computer Science Institute, Sulaimani Polytechnic University, Sulaymaniyah, Iraq
[c]Department of Computer Science, College of Science, Charmo University, Sulaymaniyah, Iraq
[d]Ministry of Education, Kirkuk Education Department of Kurdish Studies, Hawazen Preparatory School for Girls, Kirkuk, Iraq
[e]Computer Science and Engineering Department, School of Science and Engineering, University of Kurdistan Hewler, Erbil, Iraq

Email: bryar.hassan@chu.edu.iq

**Abstract**

Nowadays, cybersecurity has grown into a more significant and difficult scientific issue. The recognition of threats and attacks meant for knowledge and safety on the internet is growing harder to detect. Since cybersecurity guarantees the privacy and security of data sent via the Internet, it is essential, while also providing protection against malicious attacks. Encrypt has grown into an answer that has become an essential element of information security systems. To ensure the security of shared data, including text, images, or videos, it is essential to employ various methods and strategies. This study delves into the prevalent cryptographic methods and algorithms utilized for prevention and stream encryption, examining their encoding techniques such as advanced encryption standard (AES), Blowfish, Twofish, Salsa20, and ChaCha20. The primary objective of this research is to identify the optimal times and throughputs (speeds) for data encryption and decryption processes. The methodology of this study involved selecting five distinct types of images to compare the outcomes of the techniques evaluated in this research. The assessment focused on processing time and speed parameters, examining visual encoding and decoding using Java as the primary platform. A comparative analysis of several symmetric key ciphers was performed, focusing on handling large datasets. Despite this limitation, comparing different images helped evaluate the techniques' novelty. The results showed that ChaCha20 had the best average time for both encryption and decryption, being over 50% faster than some other algorithms. However, the Twofish algorithm had lower throughput during testing. The paper concludes with findings and suggestions for future improvements.

**Keywords:** Encoding; Decoding, Blowfish, Twofish, Salsa20, Chach20, AES.

**Introduction**

Nowadays, studying cybersecurity has grown increasingly significant. It has become increasingly harder to recognize dangers and assaults meant for knowledge and security on the internet. Cybersecurity requires urgent attention. as it assures the security and integrity of data transmission over the Internet, while also offering protection against harmful assaults. Fortunately, encryption has grown from an issue to a demand in information security systems. To protect shared data, several strategies must be implemented. The times of encryption and decryption, and throughputs (speeds) of some of the most popular block and stream encryption techniques and algorithms have been investigated in this study [1].

Moreover, cybersecurity has evolved become an increasingly significant and demanding field of research. Very grow increasingly harder to identify threats and attacks aimed at information security via the Internet. Because it enables protection against malicious attacks and guarantees the confidentiality and authenticity of data transferred via the Internet. While creating and developing web applications and websites, challenges may arise due to open-source platforms like WordPress [2], which can inadvertently expose vulnerabilities to malware attacks through images generated by such platforms. Consequently, developers employ various techniques, such as encryption, to precaution these applications against intruder attacks and preserve their integrity amidst potential intimidations and threats.

Encryption has evolved into a premier solution, emerging as a vital component within systems to safeguard sensitive information. Safeguarding information exchanged involves a variety of approaches. Utilizing the most common block and stream ciphers and algorithms, this study evaluated their encryption, decryption time, and throughput (speed) [3] advanced encryption standard (AES), Blowfish, Twofish, Salsa20 and ChaCha20 were evaluated. This study employing several sorts of photos. The current research assesses each algorithm's efficacy according to two key indicators: processing time and speed. This article explores and compares numerous symmetric key ciphers (AES, Blowfish, Twofish, Salsa20 and ChaCha20) are based on time-based encoding and decoding of picture files using Java as the programming language.

In this paper, we compare stream ciphers and block ciphers, describing the development of two stream cipher algorithms (Sales20 and ChaCha20) and three different algorithms for block ciphers: AES, Blowfish, and Twofish, as shown in figure 1. The methods used in this study, identified in cryptographic statements for cybersecurity, have been analyzed to find the best technique. In addition, the evaluation was based on outcomes related to speed and time strategies.

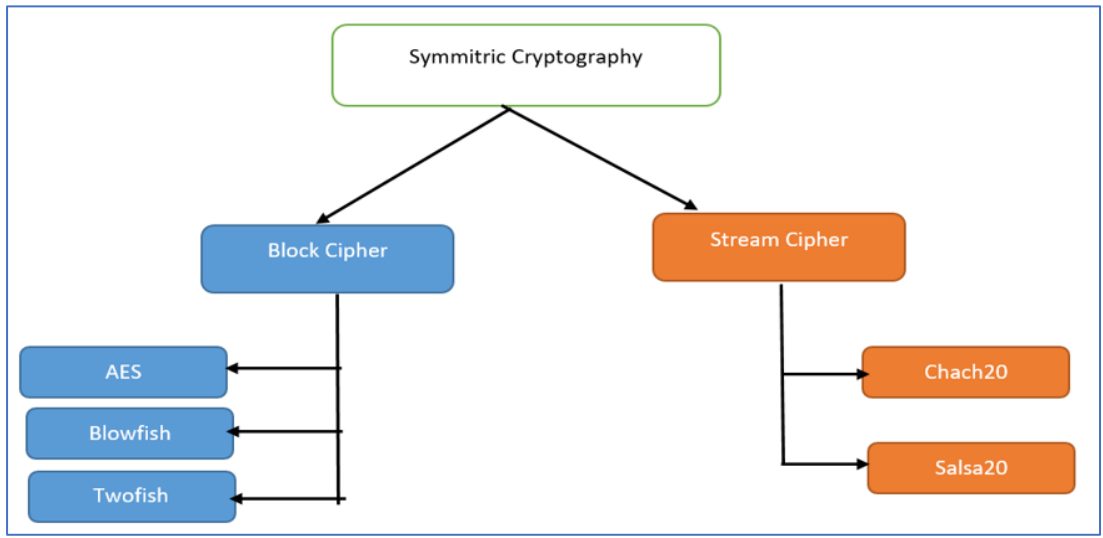

**Figure 1**: The study included block and stream cipher methods.

The primary motivation behind this research is the increasing complexity and importance of cybersecurity, particularly in the context of transmitting encrypted data. As internet threats and attacks grow more sophisticated and harder to detect, ensuring the privacy and security of online data has become a critical concern. This study highlights the significance of encryption as an essential element of information security systems, aimed at protecting data from malicious attacks. To identify the most efficient and accurate technique, it is crucial to evaluate existing methods and be prepared to enhance and develop new algorithms. This involves assessing previous techniques using selected data, then generating novel approaches to help developers identify areas for future improvement and limitations. This process will be demonstrated through statistical experiments, covering throughput during encryption and decryption, as well as the time required for cryptographic analysis. In addition, after discussing previous works, the study focuses on five algorithms: AES, Blowfish, Twofish, Salsa20, and ChaCha20. These algorithms are prevalent in cryptographic methods and were chosen for their effectiveness in stream encryption and prevention. They represent different symmetric cryptographic algorithms, including block ciphers and stream ciphers. The study also assists researchers in selecting the best key size for stream ciphers, as well as the optimal block size and key size for block ciphers [4].

Based on this motivation, the study selects plaintexts following the key size and block size standards specific to each algorithm. The selected images vary in pixel density and size to assess the statistical performance of the chosen techniques. Additionally, the study method evaluates the effectiveness of the programming language used and addresses any issues encountered during the mathematical processes involved.

However, several different objectives may influence the study. Despite these variations, certain key points consistently contribute to the research. The following outlines the main contributions of this study:

- This study examines five symmetric encryption algorithms: AES, Twofish, Blowfish, Salsa20, and ChaCha20. Performance was evaluated based on throughput and encryption time, with Java used for implementation.
- The study measured the throughput and encryption time of each algorithm using a 128-bit key, proving that ChaCha20 outperformed Twofish, Blowfish, and Salsa20 with the fastest processing time.
- Based on overall throughput, ChaCha20 and AES were identified as the most suitable algorithms for image encryption and decryption.
- The study conducted a comparative analysis of symmetric key ciphers, including key size and block size, emphasizing the importance of speed and efficiency in cryptographic techniques.

The remainder of the paper includes an overview of cyber methods, discussing how each algorithm is analyzed in cryptographic rounds. The second section reviews related works, focusing on updated reference papers. The third section includes materials for each algorithm and outlines the study's methodology as applied to the general problem. The fourth section presents the experimental results. Section five is dedicated to a detailed discussion comparing the results of this study with previous findings. Finally, the paper concludes with a discussion of conclusions and future works.

**Related works**

Recent surveys have investigated different cryptographic techniques, examining algorithms like Blowfish, Twofish, and AES within the realm of network security, cybersecurity, and cryptography [5], [6] . These studies offer an extensive exploration of encryption methods and their efficacy.

In their comprehensive analysis of Symmetric Key Cryptographic algorithms, Nema and Rizvi [7] meticulously assessed the strengths and weaknesses inherent in each algorithm. Their research findings highlighted Blowfish as surpassing its counterparts in security, adaptability, performance of encryption and memory use. Ganpati and Tyagi [8], in their assessment of Encryption Algorithms using Symmetric Keys, aimed to deepen comprehension of the cryptographic process while comparing the performance of various symmetric encryption algorithms. Their results indicated that Blowfish exhibited superior execution speeds compared to other prevalent algorithms such as Data Encryption Standard (DES), Triple Data Encryption Standard (3DES), and Advanced Encryption Standard (AES). Additionally, among the evaluated algorithms, Blowfish demonstrated superior performance metrics in terms of encryption time, decrypt time, and speed, while 3DES showed a poor performance.

Singh et al.'s [9] comparison of AES, DES, 3DES, and Blowfish algorithms concentrated on evaluating security and power consumption. According to their simulation outcomes, AES emerged as the algorithm displaying superior overall performance compared to the other algorithms scrutinized.

In 2013, Ramesh and Suruliandi [10] conducted an evaluation of specific symmetric algorithms, focusing on their effectiveness. Their experimental observations and analysis of text file sizes led them to assert that the Blowfish method exhibited superior throughput, consuming less space in memory and time for execution. When compared to DES and AES, Blowfish operated approximately four times more quickly. Additionally, in terms of memory utilization, Blowfish showed lower requirements. AES, requiring more computational resources than other algorithms, delivered comparatively inferior performance outcomes. Not only does Blowfish emerge as the swiftest encryption algorithm, but its substantial key size also contributes to its robust security profile, Therefore, it can be applied to a wide range of tasks, such as Packet encryption, online security, and random bit generation.

In a survey conducted by Yegireddi and Kumar [11] Although AES and Blowfish have flexible key structures, they are the only algorithms that provide speed and security when compared to more popular traditional encryption approaches. Extensive research on both symmetric and asymmetric cryptography algorithms has resulted in numerous studies aimed at enhancing and improving them [5], [12], [13]. Table 1 specifically highlights selected works focusing on these algorithms, particularly those relevant to the problem under study.

**Table 1**: Differences in selected symmetric cryptography Algorithm utilized in previous works.

| No. | Title of Study | Name of Cipher Algorithm | Varieties of Data Employed | Published year | Ref. |
|---|---|---|---|---|---|
| 1 | Advanced Encryption Standard (AES) Algorithm to Encrypt and Decrypt Data | AES | Text Data Cipher | 2017 | [14] |
| 2 | Novel Hybrid Encryption Algorithm Based on AES, RSA, and Twofish for Bluetooth Encryption | RSA (Rivest-Shamir-Adleman), AES, and TwoFish | Text, Image, and Video Cipher (Blutouth Cipher) | 2018 | [15] |
| 3 | Modified Advanced Encryption Standard Algorithm for Information Security | AES | Text, Image, and Video Cipher | 2019 | [16] |
| 4 | Comparison of Encryption Algorithms: AES, Blowfish and | AES, Blowfish and Twofish | Text, Image, and Video Cipher | 2020 | [17] |

| | Twofish for Security of Wireless Networks | | | | |
|---|---|---|---|---|---|
| 5 | Encrypt Medical Image using CSalsa20 Stream Algorithm | Salsa20 | Image Cipher | 2020 | [18] |
| 6 | Improving Salsa20 Stream Cipher Using Random Chaotic Maps | Salsa20 | Text, Gray Images, and Colored image cipher | 2020 | [19] |
| 7 | Comparative study on blowfish and Twofish algorithms for image encryption and decryption | Blowfish and Twofish | Image Cipher | 2020 | [20] |
| 8 | Improved Twofish Algorithm: A Digital Image Enciphering Application | Twofish | Image Cipher | 2021 | [21] |
| 9 | Rotational Cryptanalysis on ChaCha20 Stream Cipher | ChaCha20 | Text, Image and Video Cipher | 2022 | [22] |

**Methods and Materials**

The study involves analyzing test images of various sizes, discussed in this section according to different algorithms, clarified in each subsection. Following that, the methods used in cryptographic analysis are provided. For further clarification, all methods are explained with pseudocode and examples, along with the programs used in the problem statements.

*1.1. Advanced Encryption Standard*

The Advanced Encryption Standard commonly known as AES encryption, is a symmetric block cipher designed to secure data blocks of 128 bits. Following a rigorous five-year public competition, The United States government's latest encryption standard, AES, was accepted by the National Institute of Standards and Technology (NIST) in 2000. AES offers a choice of three key sizes: 128 bits, 192 bits, and 256 bits. The number of encryption rounds is determined by the key size, with 10 rounds for 128-bit keys, 12 rounds for 192-bit keys, and 14 rounds for 256-bit keys. Widely recognized as one of the most robust encryption algorithms available, AES plays a pivotal role in safeguarding sensitive data in a vast array of applications, encompassing government communications, financial transactions, and electronic medical records [17], [23], [24]. The AES method is elucidated in figure 2, detailing the encryption process.

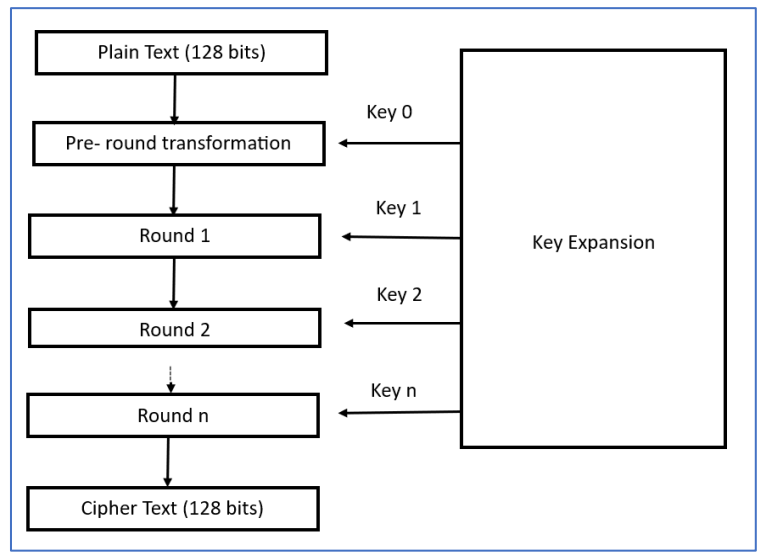

Figure 1: Structure of AES Algorithm.

### 3.2. Blowfish Algorithm

Blowfish stands out as a symmetric cipher algorithm that excels in encrypting and safeguarding sensitive data. It boasts a flexible length of the key, which can be 32–448 bits, making it well-suited for a wide spectrum of data security requirements. Blowfish's inception was driven by the need for a fast and freely available alternative to existing encryption algorithms. While it is not immune to weak key vulnerabilities, no successful attack against it has been documented. Blowfish functions in the form of symmetric block cipher using 64 bits, capable of accommodating keys of variable length. The algorithm's design is divided into two primary components: two modules: one for data encryption and one for key extension. The task of transforming a key with up to 448 bits into a number of subkey lists totaling 4168 bytes comes to the key extension function. Encrypting data is accomplished by a Feistel network with 16 rounds, ensuring thorough protection of sensitive information. Figure 3 accurately depicts the schema of the Blowfish technique. Blowfish is particularly well-suited for applications where the encryption key remains relatively static, such as secure communication links or automatic file encryption systems [1].

Furthermore, Blowfish employs a Feistel network structure, dividing input data into two halves and iteratively smearing a series of substitution and permutation operations. Despite its age, Blowfish relics popular in various applications due to its speed, robustness, and the absence of any known successful cryptanalysis.

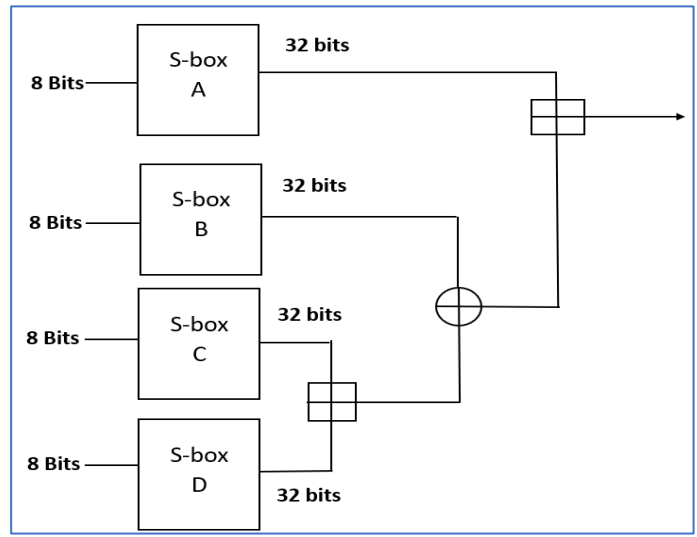

Figure 2: Blowfish encryption scheme.

### 3.3. Twofish Algorithm

A symmetric-key encryption algorithm, the Threefish cipher boasts a 128-bit block size and a flexible key length that can extend up to 256 bits. Renowned for its high level of security and adaptability, Threefish operates seamlessly on both 8-bit smart card processors and large processors. Its robust design incorporates a total of 16 rounds, ensuring that at least 32 bits of non-trivial data are processed in each round. Figure 4 provides an overview of the rounds process in accordance with the Twofish method [25].

Given two inputs, $a'$ and $b'$ are defined in equation (1) and (2), the 64-bit data is divided into halves, forming the basis for the Pseudo Hadamard Transform (PHT) [26]. This simple operation, employed for its efficiency in software implementations, can be executed using just two opcodes on modern microprocessors.

$$a' = a + b \; Mod \; 2^{32} \quad (1)$$

$$b' = a + 2b \; Mod \; 2^{32} \quad (2)$$

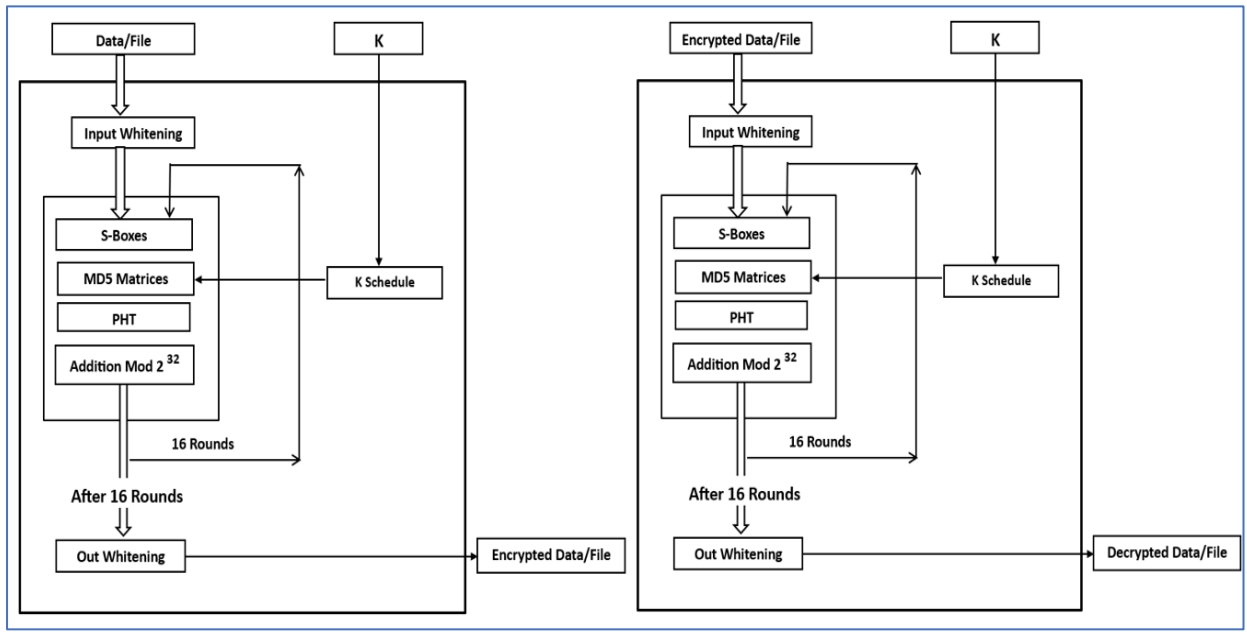

Figure 3: Encryption/decryption process.

### 3.4. Sales20 Algorithm

Salsa20 utilizes xoring the outcome of hash the private key, a nonce, and block number with the plaintext to create an encrypted a 64-byte plaintext block. With a 32-byte or 256-bit key $k = (k_0, k_1..., k_7)$ and an 8-byte or 64-bit a nonce $v = (v_0, v_1)$ as input, the stream encryption algorithm Salsa20 may operate on 32-bit or 4-byte words. The block counter $t = (t_0, t_1)$ then generates a series of keystream blocks, either 64 or 512 bits in size [27].

### 3.5. Chach20 Algorithm

256-bit keys, 64-bit nonces, and 64-bit stream positions are used by Daniel J. Bernstein's stream cipher algorithm, Salsa20, to create 512-bit keystream blocks for evaluation in the eSTREAM project. In particular, bitwise XORs, rotations, and 32-bit adds are the foundational add-rotate-xor (ARX) processes that drive its pseudorandom function. This combination of operations enables users to efficiently access any position within the keystream in constant time. However, it's important to note that certain attacks against Salsa20 have been documented.

Drawing upon stream cipher principles, the algorithm incorporates a 256-bit secret key, a 512-bit block size, and 20 rounds. Each round involves 16 XOR operations, 16 additions modulo 2^32, and 16 rotation operations (ARX operations). Additionally, the algorithm utilizes a 4x4 state matrix X, with 16 elements processed in the QuarterRoundFunction (QRF) during data encryption and decryption operations [28].

### 3.6 Implementation Method

Our implementation of various symmetric encryption algorithms in Java encompasses five algorithms: AES, Blowfish, Twofish, Salsa20, and ChaCha20. We evaluated their performance based on encryption and decryption time, throughput, and key size. The key size was set to 128 bits for AES, Blowfish, and Twofish, while ChaCha20 utilized a 256-bit key. This paper delves into a comparative analysis of these cryptosystems using the following metrics and the throughput measurement metric is considered in formula (3):

A. Encryption Time: Encryption time represents the duration required to convert plaintext into ciphertext. It is influenced by factors such as key size, plaintext block size, and encryption mode. We measured the encryption time in milliseconds during our experiments. System efficiency is strongly impacted by encryption time; faster and more responsive systems usually correspond with shorter encryption times.

B. Decryption Time: Decryption time refers to the time needed to recover plaintext from ciphertext. Similar to encryption time, shorter decryption times are desirable to maintain system responsiveness and efficiency. In our experiments, decryption time was also measured in milliseconds. Decryption time plays a crucial role in overall system performance.

C. Throughput: Throughput quantifies the speed of an encryption or method of decoding. It is computed using the entire value of the encrypted data. or decrypted per unit time, typically expressed in bits per second. Throughput is a critical metric for assessing the efficiency of encryption algorithms.

$$Throughput = \frac{Tp}{Et} \qquad (3)$$

When $Tp$ represents the total plaintext measured in kilobytes, and $Et$ denotes the encryption time in seconds.

As mentioned in the previous sections, finding the throughput and mathematical complexity for all algorithms follows a similar process. However, the functionality of each algorithm differs according to their specific standards. The analysis standards depend on whether the technique uses a block cipher or a stream cipher. For example, each algorithm analyzes the program for encryption and decryption equations to determine the elapsed time for encrypting an image (using formula 4) and decrypting an image (using formula 5). The steps for analyzing the data are referenced in figure 5, which illustrates the performance steps for image encryption and decryption for all techniques used in the study.

$$Et = ENt - STt \qquad (4)$$

$$Dt = ENt - STt \qquad (5)$$

When evaluating encryption time (Et) and decryption time (Dt), Et represents the total time taken from the start time (STt) to the end time (ENt) of the image encryption process, and Dt represents the total time taken from STt to ENt of the image decryption process.

```
// measure encryption and decryption performance
//according to throuhgput and time elapsed.
    1.  (1) Measure encryption time
    2.  startTime = getCurrentTimeMillis()
    3.  encryptedBytes = encryptImage(inputFile, encryptedFile, encryptionKey, iv)
    4.  endTime = getCurrentTimeMillis()
    5.  encryptionTime generated according to equation (4)
    6.  (2) Calculate encryption throughput in KB/s
    7.  encryptionThroughputKB = calculateThroughput(encryptedBytes, encryptionTime)
    8.  (3) Measure decryption time
    9.  startTime = getCurrentTimeMillis()
    10. decryptedBytes = decryptImage(encryptedFile, decryptedFile, encryptionKey, iv)
    11. endTime = getCurrentTimeMillis()
    12. decryptionTime generated according to equation (5)
    13. (4) Calculate decryption throughput in KB/s
    14. decryptionThroughputKB = calculateThroughput(decryptedBytes, decryptionTime)
    15. (5) print performance metrics
    16. //Print Encryption Time and Decryption Time by ms
    17. //Print Encryption Throughput and Decryption Throughput by KB/s
    18. (6) Function to get the current time in milliseconds
    19. function getCurrentTimeMillis()
    20. (7) Function to encrypt the image and Function to decrypt the image
    21. function encryptImage(inputFile, encryptedFile, encryptionKey, iv)
    22. function decryptImage(encryptedFile, decryptedFile, encryptionKey, iv)
    23. //Return the number of encrypted decrypted bytes
    24. (8) Function to calculate throughput in KB/s
    25. function calculateThroughput(bytes, timeMillis)
    26. return kilobytes / seconds
    27. (9) The problem is stopped
```

**Figure 5**: Pseudocode measure encryption and decryption performance.

However, the evaluation results, elaborated upon in next sections, offer insights into the performance attributes of each algorithm across different scenarios. Based on the overview of the techniques employed in this study, table 2 categorizes the algorithms according to block size, key size, and cryptographic types, delineating the results for test study evaluations. The cypher types are classified according to block and streams in the transmission control.

**Table 2**: Algorithms settings.

| Algorithms | Block Sizes Used (bits) | Key Size Standard (in bits) | Key Size Used (in bits) | Cipher Type |
|---|---|---|---|---|
| **AES** | 128 | 128, 192, 256 | 128 | Block |
| **Blowfish** | 64 | 32 to 448 | 128 | Block |
| **Twofish** | 128 | 128, 192, 256 | 128 | Block |
| **Salsa20** | 64 | 128, 256 | 128 | Stream |

| | ChaCha20 | 64 | 128, 256 | 128 | Stream |

**Results**

After implementing each algorithm as discussed in the previous section, the focus was on throughput and the time elapsed for encrypting and decrypting images of different sizes. Following the encryption process, which was conducted on five selected images of varying sizes measured in kilobytes, the encryption times for all types are recorded in table 3, while table 4 presents the encryption throughput. The test results reveal that Twofish requires more time for encryption compared to other techniques, whereas ChaCha20 exhibits shorter encryption times than the others. However, when considering encryption throughput, Twofish demonstrates smaller throughput sizes, while ChaCha20 showcases greater throughput in the evaluation test.

Similarly, the encryption process was carried out on the same set of five selected images, varying in size and measured in kilobytes. These images were also selected for the decryption process. The decryption times for all types are documented in table 5, while table 6 displays the decryption throughput. Results indicate that Twofish, like in encryption, requires more time for decryption compared to other techniques, whereas ChaCha20 demonstrates shorter decryption times than the others. However, in terms of decryption throughput, Twofish shows smaller throughput sizes, while ChaCha20 exhibits greater throughput in the evaluation test.

**Table 3**: Encryption time testing.

| File Name | Image size ($KB$) | Encryption Time in Milliseconds | | | | |
|---|---|---|---|---|---|---|
| | | AES | Blowfish | Twofish | Salsa20 | ChaCha20 |
| Image01.jpg | 137 | 2 | 3 | 40 | 9 | 2 |
| Image02.jpg | 795 | 13 | 15 | 252 | 13 | 5 |
| Image03.jpg | 3901 | 45 | 54 | 1223 | 50 | 39 |
| Image04.jpg | 7903 | 87 | 90 | 2305 | 168 | 73 |
| Image05.jpg | 9328 | 108 | 111 | 2771 | 125 | 80 |
| **Average** | | **51** | **54.6** | **1318.2** | **73** | **39.8** |

**Table 4**: Encryption throughput testing.

| File Name | Image size ($kB$) | Throughput for Encryption in K$B/s$ | | | | |
|---|---|---|---|---|---|---|
| | | AES | Blowfish | Twofish | Salsa20 | ChaCha20 |
| Image01.jpg | 137 | 68781.25 | 45851 | 3438.84 | 15283.74 | 68776.85 |

| File Name | Image size (kB) | | | | | |
|---|---|---|---|---|---|---|
| Image02.jpg | 795 | 61197.11 | 53037.5 | 3156.97 | 61196.81 | 159111.71 |
| Image03.jpg | 3901 | 86673.95 | 72228.15 | 3189.13 | 78006.32 | 100008.11 |
| Image04.jpg | 7903 | 90829.92 | 87802.25 | 3428.28 | 47036.90 | 108249.31 |
| Image05.jpg | 9328 | 86369.06 | 84034.76 | 3366.24 | 74622.85 | 116598.20 |
| Average | | 78770.258 | 68590.732 | 3315.892 | 55229.324 | 110548.836 |

Table 5: Decryption time testing.

| File Name | Image size (kB) | Decryption Time in Milliseconds | | | | |
|---|---|---|---|---|---|---|
| | | AES | Blowfish | Twofish | Salsa20 | ChaCha20 |
| Image01.jpg | 137 | 5 | 3 | 55 | 2 | 2 |
| Image02.jpg | 795 | 15 | 10 | 242 | 10 | 11 |
| Image03.jpg | 3901 | 131 | 49 | 1226 | 50 | 34 |
| Image04.jpg | 7903 | 158 | 95 | 2316 | 178 | 65 |
| Image05.jpg | 9328 | 120 | 114 | 2749 | 125 | 80 |
| Average | | 85.8 | 54.2 | 1317.6 | 73 | 38.4 |

Table 6: Decryption throughput testing.

| File Name | Image size (kB) | Throughput for Decryption in $kB/s$ | | | | |
|---|---|---|---|---|---|---|
| | | AES | Blowfish | Twofish | Salsa20 | ChaCha20 |
| Image01.jpg | 137 | 27510.74 | 45851 | 2500.97 | 68776.85 | 68776.85 |
| Image02.jpg | 795 | 53037.23 | 79555.85 | 3287.43 | 79555.85 | 72323.50 |
| Image03.jpg | 3901 | 29773.40 | 79598.29 | 3181.33 | 78006.32 | 114715.18 |
| Image04.jpg | 7903 | 50013.92 | 83181.05 | 3412.00 | 44394.38 | 121572.31 |
| Image05.jpg | 9328 | 77732.13 | 81823.30 | 3393.18 | 74622.85 | 116598.20 |
| Average | | 47613.484 | 74001.898 | 3154.982 | 69071.25 | 98797.208 |

Despite the varying sizes and shapes of the images, each algorithm produces different outputs. For example, Figure 6 illustrates the successful encryption and decryption process using the Twofish technique. This figure shows the image before and after encryption, highlighting the effectiveness of the algorithms within the specified period.

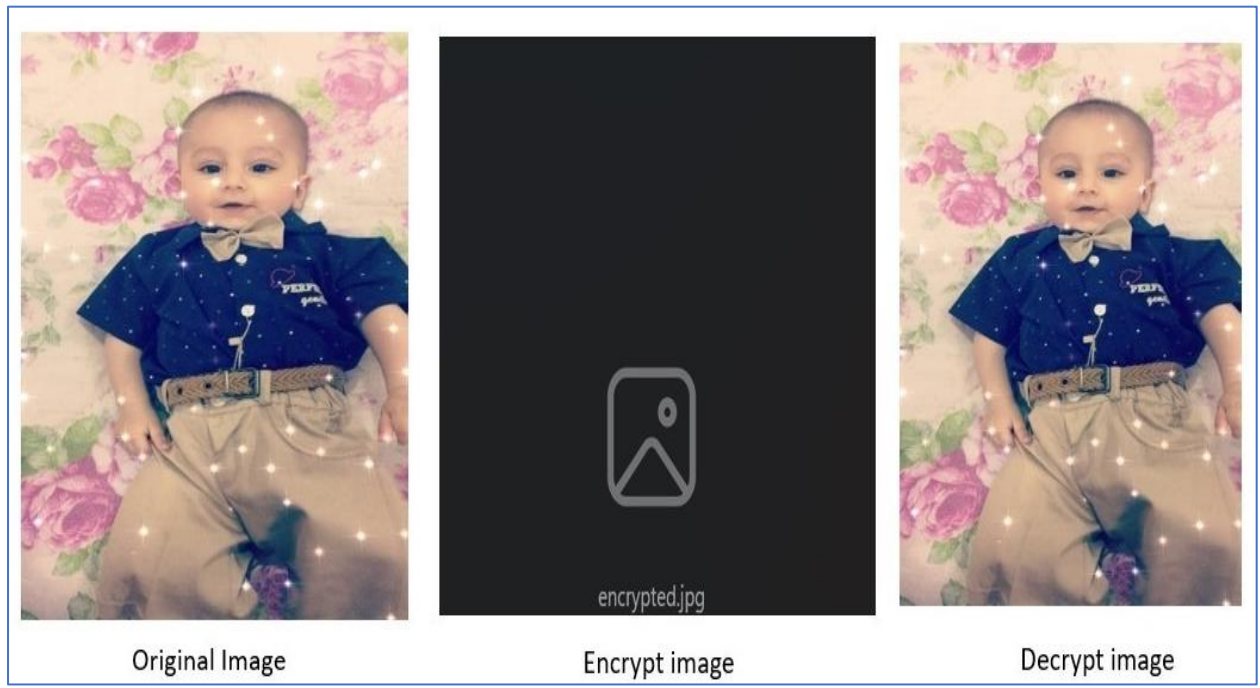

**Figure 6**: An example of encrypting/decrypting an image using the Twofish algorithm.

Figure 7 illustrate the encryption and decryption, decoding, and processing speeds as depicted in Tables 2 and 3. In terms of speed and processing time, AES shows relatively consistent performance in both encryption and decryption, with moderate times compared to other algorithms. Also, Blowfish performs similarly to AES in decryption but slightly slower in encryption and Twofish exhibits significantly longer encryption and decryption times compared to other algorithms. Formerly, Salsa20 and ChaCha20 demonstrate efficient encryption and decryption, with ChaCha20 showing the fastest average times overall.

Figure 8 also presents the dedicated throughput for both encryption and decryption. The analysis figure highlights the varying throughput performance of each algorithm, with ChaCha20 leading in both encryption and decryption tasks. ChaCha20 demonstrates the highest average throughput for both encryption and decryption, indicating its efficiency in processing data. Also, Blowfish shows the highest throughput for decryption, while AES exhibits the highest throughput for encryption. Accordingly, Twofish consistently demonstrates lower throughput values compared to other algorithms, indicating relatively slower data processing capabilities.

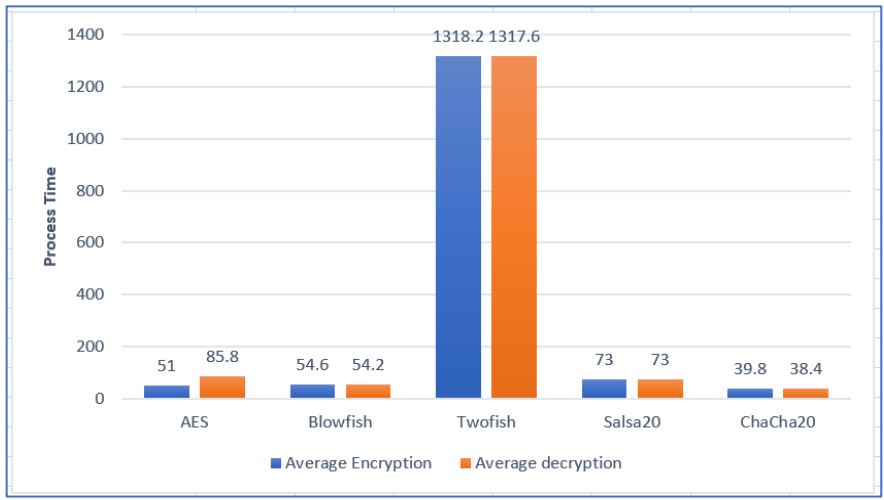

Figure 7: Time encryption/decryption process.

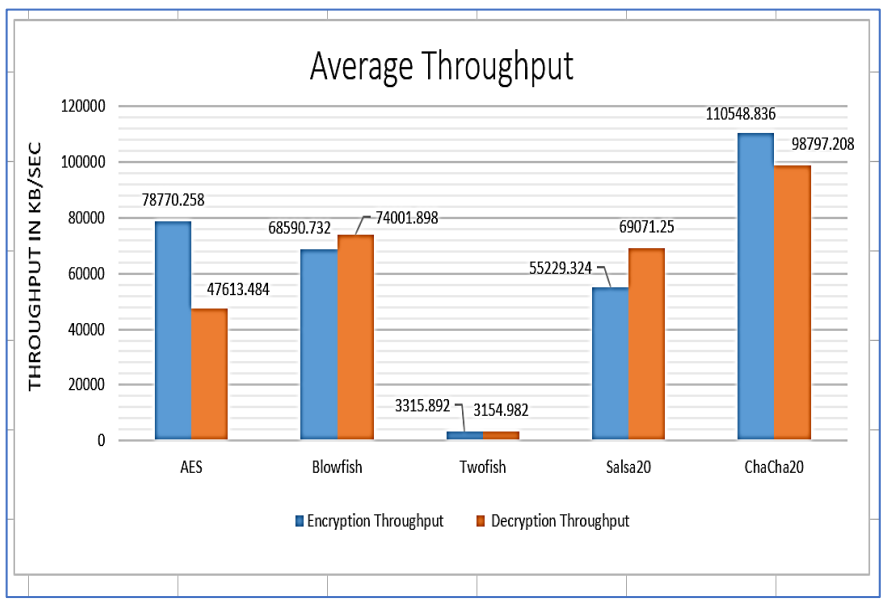

Figure 8: Throughput encryption/decryption process.

**Discussion**

ChaCha20 demonstrates the shortest encryption time, averaging 38.9 milliseconds, closely followed by AES at 51 milliseconds. Blowfish and Salsa20 comes next with an average encryption time of 54.6 milliseconds and 73 milliseconds, respectively. Twofish show longer encryption times, with 1318.2 milliseconds. These findings illustrate that the time consumption for the ChaCha20 technique is superior, consistent with previous studies across various data types, affirming its quality performance.

To further analyse and improve our results, we incorporated additional evaluation metrics and expanded the range of experimental data. The results of the AES algorithm were compared with those from Kataria and his coauthors' study to enhance our findings based on time-elapsed performance [29]. Although the image sizes vary, the comparison is based on the percentage of average time relative to the average size of images for all encrypted and decrypted images, as mentioned in related work [7], [30]. This study found that encryption required less time than decryption, contrary to other results where encryption took more time. When comparing the two cases, this study demonstrated superior evaluation and performance, achieving 1.2% for encryption and 1.9% for decryption in terms of average time per average size. Conversely, Kataria and his coauthors had higher averages [29]. Furthermore, researchers can compare these results with other specific cryptographic algorithms for further evaluation. For further clarification, Table 7 demonstrates superior performance in presenting specific results. Additionally, this work can be enhanced using statistical non-parametric tests to evaluate the problem and its solutions. This approach, similar to some studies that optimize problems based on heuristic results, can help in assessing and addressing the impacts effectively [31].

Table 7: Comparison of time-efficient encryption and decryption using the AES algorithm for current and previous study.

| File Name | Result of This Study | | | Kataria's Result [29] | | |
|---|---|---|---|---|---|---|
| | Image size ($kB$) | Encryption Time | Decryption Time | Image size ($kB$) | Encryption Time | Decryption Time |
| **Image01.jpg** | 137 | 2 | 5 | 153 | 1600 | 1100 |
| **Image02.jpg** | 795 | 13 | 15 | 118 | 1700 | 1200 |
| **Image03.jpg** | 3901 | 45 | 131 | 196 | 1700 | 1240 |
| **Image04.jpg** | 7903 | 87 | 158 | 868 | 2000 | 1200 |
| **Image05.jpg** | 9328 | 108 | 120 | 312 | 1800 | 1300 |
| **Average** | 4412.8 | **51** | **85.8** | 329.4 | 1760 | 1208 |
| **Average time per size** | | 0.0116 | 0.0194 | | 5.343 | 3.667 |
| **Parentage average time per average size** | | %1.2 | %1.9 | | %534.30 | %366.73 |

The analysis investigates into the nuanced performance attributes of various encryption algorithms, considering both encryption period and throughput metrics. Notably, ChaCha20 emerges as a standout performer, showcasing remarkable efficiency in terms of throughput. Its efficient design and enhanced operations contribute to its ability to handle data rapidly and effectually. Equally, AES illustrates in terms of rapid encryption times, leveraging its progressive cryptographic techniques to expedite the encryption process without compromising security. However, Twofish presents a different profile, considered by especially longer encryption times when associated to its counterparts. This extended process duration proposes a trade-off between security and speed, highlighting the position of selecting the most appropriate encryption algorithm based on detailed application requirements. Besides, to delve deeper into cybersecurity algorithms, it is necessary to pinpoint precise solutions. This involves optimizing real-world applications, particularly those emphasizing security, like the Integrated Cyber-Physical-Attack for Manufacturing System [32]. Such advancements will play a crucial role in verifying the integrity of these techniques in future endeavors.

**Conclusions**

Recently, with the internet and network services expanding at a rapid rate, encryption techniques are crucial for protecting confidentiality of data. The current study examined the five most symmetric encryption algorithms namely: AES, Twofish, Blowfish, Salsa20 and Chach20. The measurements used for assessment of performance are throughput and encrypt time. The prototype has been written in Java and compiled using the JDK 7.1 programming kit using the NetBeans IDE7.1.2 and the default configurations and an encryption key bit size of 128 was used. The outcomes of the studies demonstrate that the Chacha20, works faster than the Twofish, Blowfish, and Salsa20 strategies and it has indicated a shorter time to process. The Chacha20 and AES strategy is more appropriate for image decryption and encryption based to the overall findings according to the throughput detail. . For future reading, the authors advise the reader could optionally read the following research works [33][34][35][36][37][38][39][40][41][42][43][44].

**Data availability:** Data will be made available on request.

**Conflicts of interest:** The authors declare that they have no known competing financial interests or personal relationships that could have appeared to influence the work reported in this paper.

**Funding:** The authors did not receive support from any organization for the submitted work.